\documentclass[%
 reprint,
 amsmath,amssymb,
 aps,
 prl,
 superscriptaddress
]{revtex4-2}
\usepackage[utf8]{inputenc}

\usepackage{graphicx,xcolor}
\usepackage{dcolumn}
\usepackage{bm}
\usepackage[percent]{overpic}
\usepackage{physics}
\usepackage{mathtools}

\definecolor{darkGreen}{RGB}{0,110,0}
\definecolor{darkBlue}{RGB}{0,0,130}
\usepackage[colorlinks,citecolor=darkGreen,linkcolor=darkBlue,%
urlcolor=blue,hyperindex]{hyperref}

\usepackage{ulem} 

\newcommand{\be}{\begin{equation}}
\newcommand{\ee}{\end{equation}}

\newcommand{\diag}{\mathbf{diag}}
\renewcommand{\(}{\left(}
\renewcommand{\(}{\left(}
\renewcommand{\)}{\right)}
\renewcommand{\[}{\left[}
\renewcommand{\]}{\right]}

\usepackage{bm}

\begin{document}

\title{Finite-temperature quantum discordant criticality}

\author{Poetri Sonya Tarabunga}
\affiliation{The Abdus Salam International Centre for Theoretical Physics, strada Costiera 11, 34151 Trieste, Italy}
\affiliation{SISSA, via Bonomea, 265, 34136 Trieste, Italy}
\author{Tiago Mendes-Santos}

\affiliation{Max-Planck-Institut für Physik komplexer Systeme, 01187 Dresden, Germany}
\affiliation{The Abdus Salam International Centre for Theoretical Physics, strada Costiera 11, 34151 Trieste, Italy}

\author{Fabrizio Illuminati}
\affiliation{Dipartimento di Ingegneria Industriale, Universit\`a degli Studi di Salerno,
Via Giovanni Paolo II, 132 I-84084 Fisciano (SA), Italy}

\author{Marcello Dalmonte}
\affiliation{The Abdus Salam International Centre for Theoretical Physics, strada Costiera 11, 34151 Trieste, Italy}
\affiliation{SISSA, via Bonomea, 265, 34136 Trieste, Italy}
\begin{abstract}

In quantum statistical mechanics, finite-temperature phase transitions are typically governed by classical field theories. In this context, the role of quantum correlations is unclear: recent contributions have shown how entanglement is typically very short-ranged, and thus uninformative about long-ranged critical correlations. In this work, we show the existence of finite-temperature phase transitions where a broader form of quantum correlation than entanglement, the entropic quantum discord, can display genuine signatures of critical behavior. We consider integrable bosonic field theories in both two- and three-dimensional lattices, and show how the two-mode Gaussian discord decays algebraically with the distance even in cases where the entanglement negativity vanishes beyond nearest-neighbor separations. Systematically approaching the zero-temperature limit allows us to connect discord to entanglement, drawing a generic picture of quantum correlations and critical behavior that naturally describes the transition between entangled and discordant quantum matter. 

\end{abstract}

\maketitle

\paragraph*{Introduction. -}
Is there any genuine quantum mechanical effect characterizing phase transitions at finite temperature? According to Ginzburg-Landau theory, such transitions are governed by thermal fluctuations, and the related classical critical exponents: the latter predict the behavior of all thermodynamic variables to be dictated solely by the corresponding classical universality class. The fate of quantum correlations at thermal critical points is considerably less understood. Quantum correlations are not straightforwardly bound by the above argument: in particular, the latter is still compatible with long-ranged, quantum mechanical correlations, as long as they do not compromise the correct scaling of thermodynamic variables.

A recent set of works has initiated the investigation of entanglement, a most prominent form of nonlocal quantum correlation~\cite{Amico2008,l-15}, at thermal phase transitions in a broad range of models \cite{Lu_2019_2, Lu2019, Wald_2020, Wu_2020}. Exploiting entanglement negativity~\cite{vw-02} and its Renyi modification~\cite{cct-12} - respectively a computationally convenient entanglement monotone and its finite-temperature proxy -, these studies have consistently supported the fact that, while entanglement can indeed be finite at short range, it inevitably dies out exponentially fast at long-distances. In particular, while negativities between neighboring degrees of freedom could still be sensitive to phase transitions (since they are related to expectation values of local operators), this will be due to the presence of local entanglement at the boundary between partitions, and thus not related to long-distance physics. The corresponding physical picture thus supports the fact that finite-temperature transitions do not host long-range quantum correlations in the form of entanglement related to the violation of the Peres criterion for separability~\cite{peres1996separability,HORODECKI1997333}.

\begin{figure}
    \centering
    \includegraphics[width=0.48\textwidth]{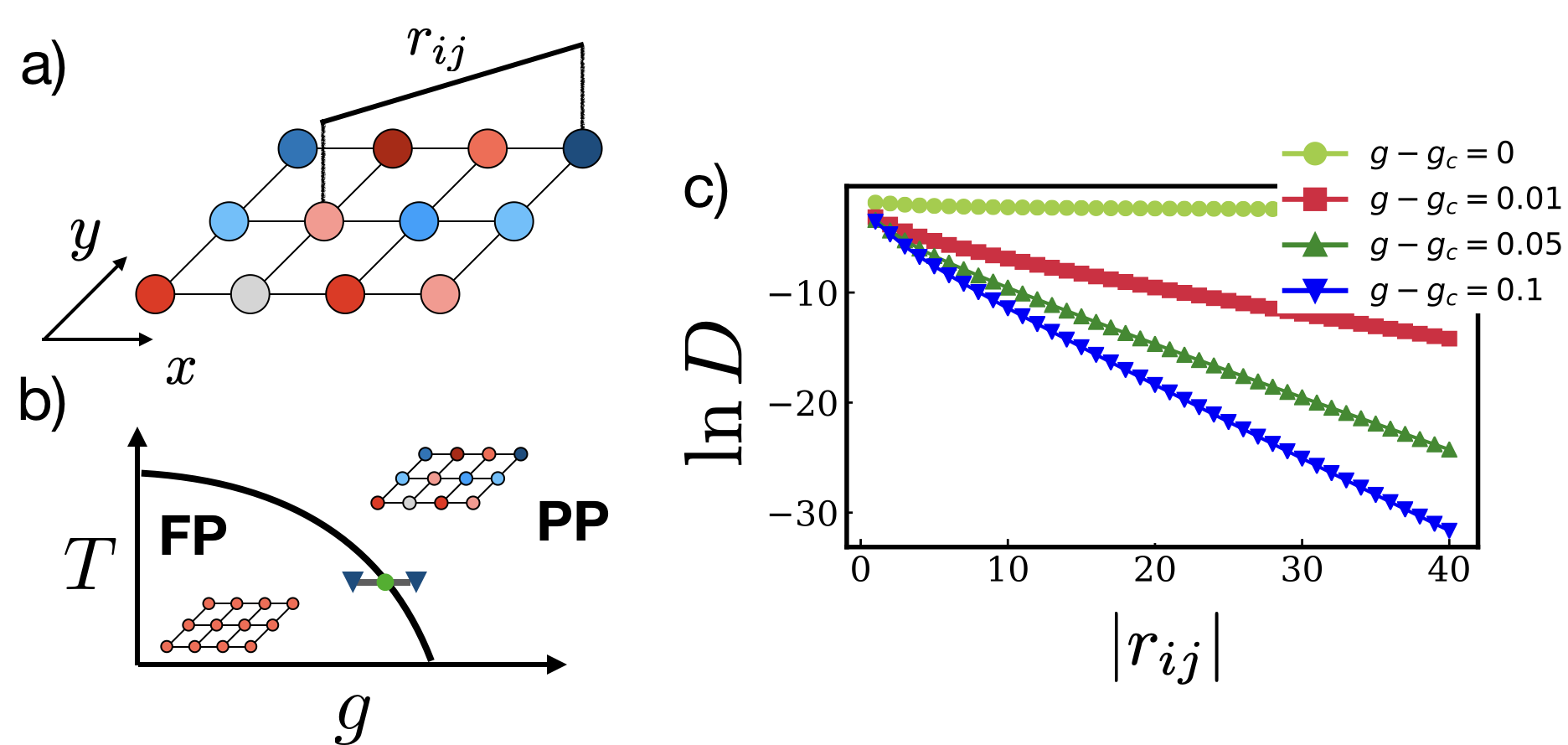}
    \caption{Quantum discordant criticality in two-dimensional systems. Panel (a): schematics of the model. Gaussian bosonic variables are defined on a square lattice: we are interested in the quantum correlations between pairs of sites $i,j$ at distance $r_{ij}$. Panel b): schematics of the finite temperature phase diagram of the model in Eq.~\eqref{eq:H}~\cite{Lu2019}, hosting a ferromagnetic (FP) and a paramagnetic phase (PP). Panel (c): decay of the entropic quantum discord along a cut at $T=XXXX$ (gray line in (b)): the decay is exponential away from the critical line, $g\neq g_c$, while it is algebraic along it.}
    \label{fig:Fig1}
\end{figure}

In this work, we show that a more basic form of quantum correlation, the entropic quantum discord (EQD), which is defined on all quantum states, is typically nonvanishing even on separable states, and reduces to entanglement on pure states \cite{Ollivier_2001, Henderson_2001}, can display genuine critical behavior at finite temperature. We consider free bosonic Gaussian theories in two- and three-dimensional hypercubic lattices~\cite{Lu2019}, both of them undergoing finite-temperature transitions from an ordered to a disordered phase (see Fig.\ref{fig:Fig1}a-b), described by the mean-field Ising universality class. In both models, entanglement between two bosonic modes is short-ranged at finite temperature: indeed, one observes sudden death of entanglement~\cite{Acin2008,Sherman_2016,Hart_2018,Wald_2020}, as the negativity vanishes beyond nearest-neighbors. Quantum discord is short-ranged both in the low-$T$ and high-$T$ phases: however, at finite-temperature critical points, such quantum correlations become quasi-long-ranged. In both 2D and 3D, the EQD between distant bosonic modes displays a power-law decay, a hallmark of critical behavior (see Fig.\ref{fig:Fig1}c). In the 2D case, we also observe that the EQD obeys universal scaling collapse, which is governed by the critical exponent $\nu$ - signalling that, remarkably, such genuine quantum correlations are governed by the same critical exponent as the correlation length. We deem this framework quantum discordant criticality (QDC).

Before embarking in a detailed description of the results, it is useful to provide a qualitative picture of our findings. Differently from entanglement, the EQD characterizes the degree of incompatibility between classical and quantum correlation functions, irrespective of separability criteria. This implies that, while at a finite-temperature critical point one can typically describe the state of two distant degrees of freedom $A$ and $B$ by a separable (i.e. unentangled) density matrix, yet the latter can display a genuine quantum character: at odds with classical correlations, the mutual information obtained by a collective measurement of the global system $AB$ is different from the one obtained by measuring subsystems $A$ and $B$ in sequence. Quantum discordant criticality thus emphasizes the role of sensitivity to measurements, rather than that of separability (which is instead pivotal in entanglement-driven quantum phase transitions), and captures aspects complementary to other general quantum features, such as quantum coherence~\cite{Roscilde2016}, that depend on the choice of the local reference basis.

\paragraph*{Entropic quantum discord. -} We first review some general properties of quantum discord, and of the models we are considering in the present work. 
In classical information theory, the amount of correlations between two classical random variables $X$ and $Y$ can be quantified by their mutual information, defined as
\be \label{eq:cmi}
I(X:Y)=H(X)+H(Y)-H(X,Y),
\ee
where $H(X)=-\sum_x p_X(x) \log p_X(x)$ is the classical Shannon entropy and $H(X,Y)$ is the joint entropy for $X$ and $Y$. Equivalently, we can express Eq.~\ref{eq:cmi} in terms of the conditional entropy $H(X|Y)$ as
\be \label{eq:cmi2}
J(X:Y)=H(X)-H(X|Y).
\ee
That the two quantities defined in Eq. \ref{eq:cmi} and Eq. \ref{eq:cmi2} are classically equivalent follows from Bayes rule, $p_{X|Y}=p_{X,Y}/p_Y$.

The situation is different for quantum systems. The quantum analogue of the mutual information for a a bipartite quantum system $AB$ can be defined as
\be
I(\rho_{AB})=S(\rho_A)+S(\rho_B)-S(\rho_{AB}),
\ee
where $S(\rho)=-\tr(\rho \log \rho)$ is the von Neumann entropy of the density matrix $\rho$. The quantum mutual information measures the total (quantum and classical) amount of correlation present in a quantum state. On the other hand, the quantum version of $J(X:Y)$ is obtained by considering the conditional state of subsystem $A$ after a measurement is performed on subsystem $B$. Let us denote by $\{\Pi_j^B\}$ a positive operator-valued measurement (POVM) describing a generalized measurement performed on $B$. The quantum conditional entropy after the measurement is given by
\be
S_{\Pi}(\rho_{A|B})=\sum_i p_i S(\rho_{A|i}),
\ee
where $p_i=\tr(\rho_{AB}\Pi_i^B)$ and $\rho_{A|i}=\tr_B(\rho_{AB}\Pi_i^B)/p_i$. The quantum version of $J(X:Y)$ then reads
\be
J_{\Pi}(\rho_{AB})=S(\rho_{A})-S_{\Pi}(\rho_{A|B}).
\ee
At variance with the classical case, in general the quantum versions of $I(\rho_{AB})$ and $J(\rho_{AB})$ are not equivalent. Such difference arises due to quantum effects, and can be exploited to measure quantum correlations in the system. This difference has been named entropic quantum discord \cite{Ollivier_2001,Henderson_2001}, and is formally defined as
\be\label{eq:disc}
\begin{split}
D(\rho_{AB}) & = I(\rho_{AB})-\max_{\{\Pi_j^B\}}J_{\Pi}(\rho_{AB}) \\
& = S(\rho_B)-S(\rho_{AB})+\min_{\{\Pi_j^B\}}S_{\Pi}(\rho_{A|B}),
\end{split}
\ee
whereas the classical correlation reads
\be \label{eq:cc}
C(\rho_{AB})  = S(\rho_{A})-\max_{\{\Pi_j^B\}}S_{\Pi}(\rho_{A|B}),
\ee
where the optimization is taken over the set of all possible POVM measurements on subsystem $B$.

Quantum discord is able to measure quantum correlations not captured by entanglement, in a sense that it can be present even in separable {\it mixed} states. The states with vanishing quantum discord are called classical-quantum states, and take the form
\be
\rho_{cq}=\sum p_i \ket{A_i}\bra{A_i} \otimes \rho_{Bi},
\ee
where $\{A_i\}$ forms an orthonormal basis for subsystem $A$ and $\rho_{Bi}$ are generic states of subsystem $B$. 

It is important to remark that the EQD is in general asymmetric in $A$ and $B$ and as such represents the weakest and at the same time the broadest element in the hierarchy of nonlocal quantum correlations; indeed, while all entangled states have necessarily nonzero discord, the opposite does not hold, as the set of classical-quantum states forms a zero-measure subset of the set of all separable mixed states \cite{Ferraro_2010}. EQD reduces to entanglement on pure states (and thus vanishes on product states); moreover, it vanishes identically on classical states, thus providing a {\it{bona fide}} measure of quantumness. 

Concerning the relation with coherence, it is immediate to see from the definitions that a bipartite state has vanishing discord if and only if the local reductions (reduced states of the subsystems) are incoherent. In other words, coherence in some local basis is a necessary and sufficient condition for a nonvanishing bipartite discord \cite{AdessoJPA2016,LesanovskyPRA2019}, reflecting the intuitive notion that classicality is intimately related to the absence of quantum superpositions in a specific local reference frame. 

The role of the EQD in many-body systems has been discussed in several contexts~\cite{DeChiara2018}, and particularly for spin systems \cite{Dillenschneider2008,Sarandy2009,Sarandy2012,Sarandy2013}. For the case of thermal phase transitions, the EQD was investigated in Refs.~\cite{Rigolin2010,Rigolin2010v2,Serra2010,Tomasello2011,Busch2013}.
The EQD exhibits a finite-temperature crossover with universal
scaling behavior in the quantum critical fan region associated with a one-dimensional quantum critical point \cite{Tomasello2011}.
Furthermore, for spin models, the scaling of the two-site EQD is similar to the one of two-body correlation functions; For instance~\cite{Huang2014}, the EQD decays polynomially in gapless (critical) ground-state systems in one dimension~\footnote{Approximated analytical expressions for the EQD can be obtained for a broad class of systems for the case of two-qubits (similarly to the case of entanglement and concurrence), which emphasize the relation between two-body EQD and correlation functions \cite{Alber2010}. However, it is worth mentioning that such expressions for the EQD are not valid for arbitrary states, and one cannot immediately determine the behavior of the EQD from two-body correlations, even for the case of two spins \cite{Adesso2011,Huang2013}. Examples in which the correlation length diverges, but the long-ranged EQD vanishes exist, as illustrated here.}. In contrast, the pairwise concurrence \cite{Amico2008} and the two-body negativity \cite{Lu2019} typically decay exponentially with the distance. 

\paragraph{Model Hamiltonian and phase diagram. -}
We consider a specific model, introduced in Ref.~\cite{Lu2019}, describing a ensemble of Gaussian bosonic variables $(\pi_{\vec{r}}, \phi_{\vec{r}})$ arranged on hypercubic lattices. 
The system Hamiltonian reads
\be \label{eq:H}
H=\frac{1}{2} \sum_{\vec{r}} \left( \pi_{\vec{r}}^2 +m^2 \phi_{\vec{r}}^2 \right) +  \frac{1}{2}  \sum_{\langle \vec{r},\vec{r}' \rangle}K \left( \phi_{\vec{r}} -\phi_{\vec{r}'}   \right) ^2,
\ee
defined on a $d$-dimensional cubic lattice of $N$ sites with periodic boundary conditions imposed on all spatial directions. The model can be regarded as a mean-field approximation of the transverse-field Ising model, where the Gaussian fluctuations are taken into account. It hosts a finite-temperature phase transition in the mean-field Ising universality class, related to the underlying $\mathbb{Z}_2$ symmetry $\phi\rightarrow -\phi $, and it features a
physical mass inversely proportional to the correlation length that obeys
\begin{align} 
    m(g)=
    \begin{cases}
    \sqrt{g-g_c} \quad &\textrm{for} \quad g>g_c \\
    \sqrt{2(g_c-g)} \quad &\textrm{for} \quad g<g_c,
    \end{cases}
\end{align}
where $g_c (T, K, m)$ is the critical point, that is a function of $K, m$ and the temperature $T$. 

The equilibrium thermal states of the model are Gaussian states, i.e. they are completely characterized by the covariance matrix $\sigma$, that can be computed analytically from the two-point correlation functions (See Ref.~\cite{Lu_2019_2,supmat}). The covariance matrix can be conveniently expressed as
\be
\sigma=
\begin{pmatrix}
a & 0 & c & 0 \\  
0 & a & 0 & d \\
c & 0 & a & 0 \\
0 & d & 0 & a \\
\end{pmatrix},
\ee
with 
$a=\sqrt{\sigma_{\phi}(0)\sigma_{\pi}(0)}, c=\sigma_{\phi}(\vec{r}-\vec{r}')\sqrt{\sigma_{\pi}(0)/\sigma_{\phi}(0)}$ and 
$d=\sigma_{\pi}(\vec{r}-\vec{r}')\sqrt{\sigma_{\phi}(0)/\sigma_{\pi}(0)}$,
where $\sigma_{\phi(\pi)}$ are two-body correlation function of the field $\phi(\pi)$. 

The entanglement properties of the model have been characterized in Ref.~\cite{Lu2019}. Short-range entanglement is still sensitive to the transition: in particular, the area law coefficient of the entanglement negativity is singular at the critical point. However, this behavior can be traced back to the fact that boundary terms are very sensitive to local correlation functions. 
Most importantly, entanglement related to violations of the positive-partial transpose criterion is insensitive to long-distance critical properties. Indeed, by considering a suitable tripartite Renyi negativity that allows to trace out such local terms, it has been shown that there is no residual long-range entanglement at the transition \cite{Lu2019}. Similarly, two-mode negativity is exponentially decaying (if not exactly vanishing) as a function of the distance between the modes for any temperature. These facts illustrate how entanglement related to violations of the positive-partial transpose criterion is short ranged, and thus, unrelated to the long-range nature of correlations at criticality. Similar results have also been observed in the context of the 2D quantum Ising model~\cite{Wu_2020}.

\paragraph{Gaussian quantum discord. - }
The evaluation of the EQD is a difficult task in general, due to the need to optimize over the set of all possible measurements. Indeed, it has been shown that computing the EQD in a generic quantum system is NP-complete \cite{huang2014-2}. On the other hand, in the case of Gaussian states, if one restricts to generalized Gaussian measurements, the corresponding Gaussian quantum discord is computable~\cite{Giorda_2010,Adesso_2010}. It has been later shown that the Gaussian discord is in fact optimal and thus coincides with the exact EQD for all two-mode Gaussian states~\cite{Pirandola2014}.

Consider then a two-mode Gaussian state $\rho_{AB}$, which is completely characterized by its covariance matrix $\sigma$ whose elements are $\sigma_{ij}=\tr(\rho_{AB}\{R_iR_j\}_+)$, where $R=(q_A, p_A, q_B, p_B)$ is the vector of the canonical quadrature operators. The covariance matrix can always be brought to a standard form by means of local unitary transformations:
\be
\sigma=
\begin{pmatrix}
\alpha & \gamma \\  
\gamma^T & \beta
\end{pmatrix},
\ee
where $\alpha=\diag\(a,a\)$, $\beta=\diag\(b,b\)$, and $\gamma=\diag\(c,d\)$. This state is fully specified by its symplectic invariants: $\tilde{A}=\det\alpha$, $\tilde{B}=\det\beta$, $\tilde{C}=\det\gamma$, and $\tilde{D}=\det\sigma$~\cite{Illuminati2004}. Due to the invariance of the discord under local unitaries, we can work with $\sigma$ in the standard form, and the discord can be expressed in terms of the four symplectic invariants. 

A Gaussian POVM measurement on mode $B$ can be described by $\Pi_B(\eta)=\pi^{-1} \hat{W}_B(\eta) \Pi^B_0 \hat{W}_B^{\dagger}(\eta)$, $\int d^2\eta \Pi^B(\eta)=1$, where $\hat{W}_B(\eta)$ is the Weyl operator and $\Pi^B_0$ is the density matrix of a pure, single-mode Gaussian state, whose covariance matrix is denoted as $\sigma_0$. After the measurement described by $\Pi_B(\eta)$, the covariance matrix of the conditional state is given by $\epsilon=\alpha-\gamma(\beta+\sigma_0^{-1})\gamma^T$, which is, remarkably, independent of the measurement outcome. Thus, the Gaussian discord can be simply written as
\be \label{eq:gqd}
D(\rho_{AB})=f\(\sqrt{\tilde{B}}\)-f(\nu_-)-f(\nu_+)+\inf_{\sigma_0} f\(\sqrt{\det\epsilon}\),
\ee
where $f(x)=(\frac{x+1}{2})\log(\frac{x+1}{2})-(\frac{x-1}{2})\log(\frac{x-1}{2})$ and the symplectic eigenvalues of $\sigma$ are given by $\nu^2_{\pm}=\frac{1}{2} \(\Delta \pm \sqrt{\Delta^2-4\tilde{D}}\)$ with $\Delta=\tilde{A}+\tilde{B}+2\tilde{C}$.

\begin{figure} [t]
    \centering
    \includegraphics[width=0.48\textwidth]{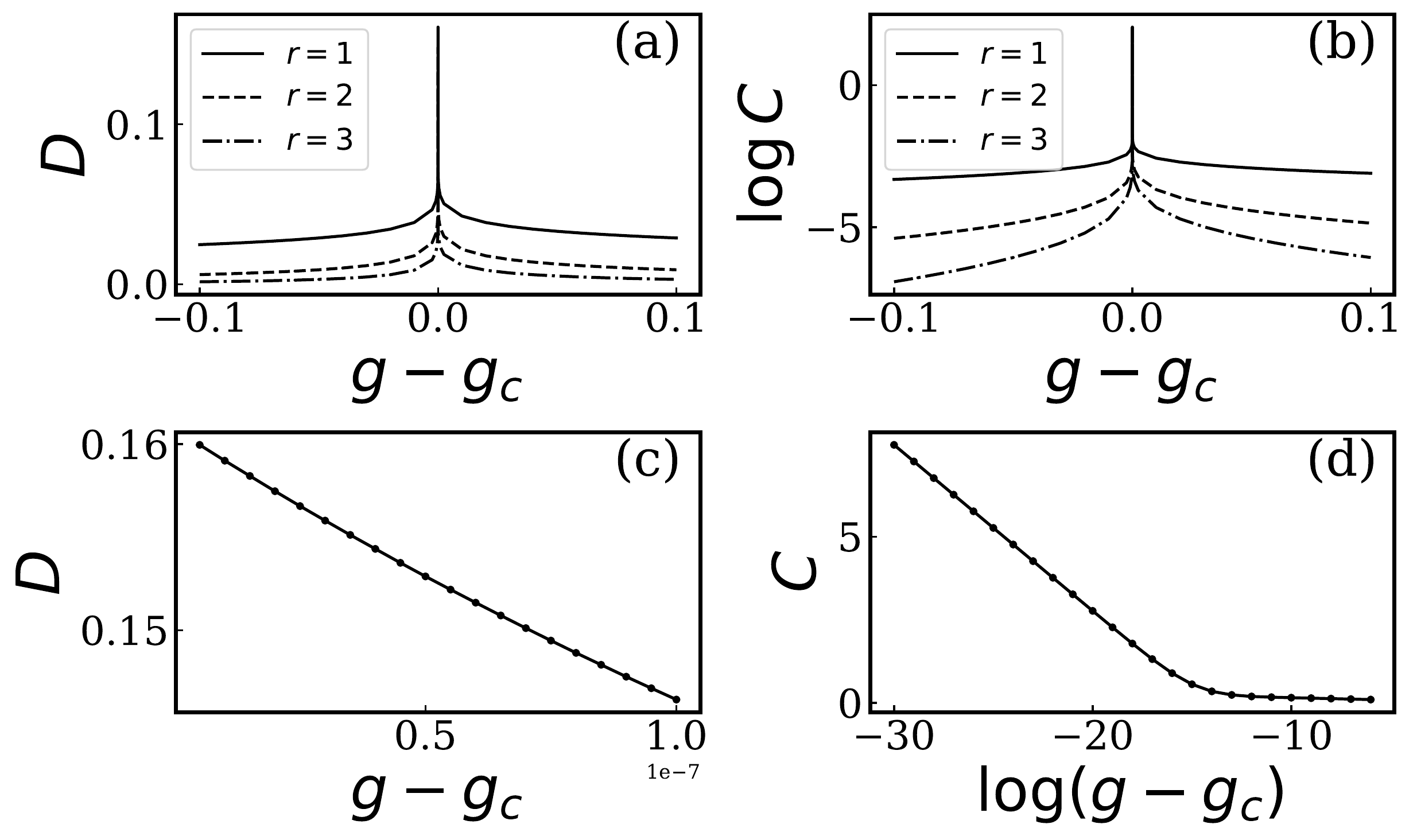}
    \caption{Behavior of (a) the quantum discord and (b) the classical correlations on a 2D lattice between two modes placed on sites at distances $r=1, 2, 3$ as a function of $g-g_c$ at $T=0.2$ for $N=1000\times1000$ sites. Behavior of the (c) quantum discord and (d) classical correlation close to the critical point at $r=1$.}
    \label{fig:singular}
\end{figure}

\paragraph{Discordant quantum criticality: short-distance properties. -} We are now in a position to characterize the phase diagram of Eq.~\eqref{eq:H} using the EQD. We will start from the 2D case, and set $K=1$ as energy unit.

We first focus on discord at fixed distance: this is a useful check to understand whether the EQD is sensitive to the same type of short-range correlations as the negativity is - a sanity check before moving to investigate the long-range behavior. In Fig.~\ref{fig:singular}a, we report the behavior of the EQD for inter-site distances $r=1,2,3$ in a system of $N=10^6$ sites: all cases are characterized by a local maximum at $g=g_c$, similarly to classical correlations (Fig.~\ref{fig:singular}b). In Fig.~\ref{fig:singular}c,d, we compare the approach to criticality of the quantum discord $D$ (Eq. \ref{eq:gqd})and the classical correlations $C$ (Eq. \ref{eq:cc}) at $r=1$: while classical correlations diverge, the EQD remains finite, scaling as $D \sim |t|$, where $|t|$ is either $g-g_c$ or $T-Tc$. This signals the fact that the latter carries qualitatively different information with respect to classical correlations only. This is confirmed further by the finite-size scaling collapse~\cite{supmat}, that is reminiscent of the one found for the negativity in Ref.~\cite{Lu2019}.

\paragraph{Discordant quantum criticality: long-distance properties. -} Oppositely to what happens at short distances, the long-distance decay of the EQD is fundamentally distinct from that of the entanglement negativity. 
In Fig.~\ref{fig:Fig1}c, we show the decay of $D$ versus distance for several values of $g$ for the 2D model. 

\begin{figure}[t]
    \centering
    \includegraphics[width=0.5\textwidth]{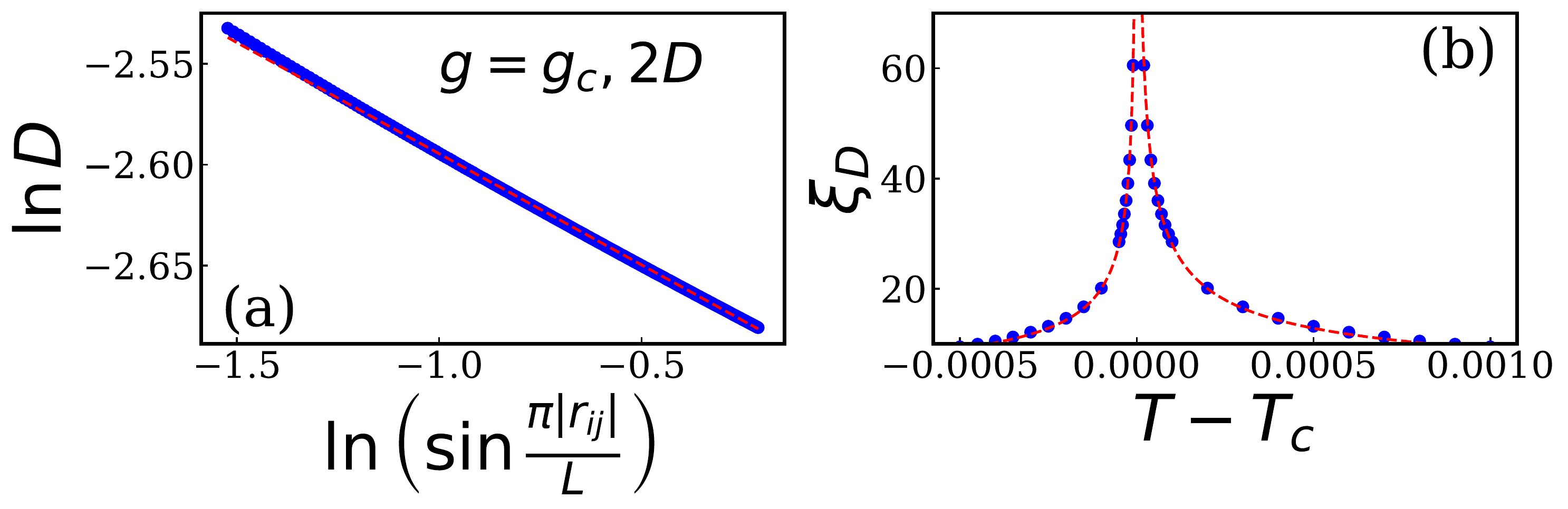}
    \caption{(a) Decay of the entropic quantum discord on a 2D lattice at the critical point. (b) Behavior of discord length $\xi_D$ near the critical point for $g(T_c=0.4)$. The red dashed line is the power law fit $\xi_D=a_{\pm}|T-T_c|^{-\nu_D}$. We find the power law exponent to be $\nu_D=0.487 \pm 0.017$, close to $\nu=1/2$ for the classical correlation length in the mean-field Ising universality class.}
    \label{fig:qd_scaling}
\end{figure}

Away from the critical point, the EQD vanishes exponentially with the distance: $D(r) \sim e^{-r/\xi_D}$, where we have introduced the {\it{discord length}} $\xi_D$ which plays a similar role as the physical correlation length. We see that the discord length is diverging as we move towards the critical point, similarly to the physical correlation length. At the critical point, both 2D and 3D models display a characteristic power law decay, as shown in Fig.~\ref{fig:qd_scaling}. In the same range of parameters, the negativity vanishes beyond a few sites. We point out that, in addition to the EQD, the geometric discord, defined as a suitable distance from the set of classical-quantum states, also displays similar signatures of the critical behavior~\cite{supmat}.

We find worth commenting now on the potential generality of quantum discordant criticality (QDC). Based on our results and on general considerations applicable also to spin-1/2 systems~\cite{Huang2014,Roscilde2016}, we envision two likely scenarios. In the first instance, QDC is a specific property of certain finite-temperature transitions. In the second one, QDC is widespread and in fact applicable to all finite-temperature transitions. Both scenarios require investigations going beyond classical field theory, calling for the development of a thermal field theory for phase transitions of quantum models at finite temperature (even if, remarkably, quantum correlations are still dictated by classically predictable critical exponents). The second scenario may potentially arise naturally in spin-1/2 systems (and thus extend to the 2D Ising model), the only case where the quantum discord - like entanglement - is directly tied to correlation functions: for instance, the EQD has been show to decay algebraically for critical phases in the 2D $XY$ model~\cite{Roscilde2016}. While, from a many-body viewpoint, this second scenario would make the phenomenology we describe somehow less interesting with respect to the first one (even if it would remarkably enable an unbiased method to distinguish a purely classical transition from a finite-temperature transition in a quantum system), it would anyway uncover an unexpected, generic setting where to investigate the quantum-to-classical crossover directly at the many-body level, under very minimal assumptions.

\paragraph{Conclusions and outlook. -} We have reported evidence of genuine, operatively meaningful quantum correlations at thermal quantum phase transitions.
In the context of Gaussian bosonic theories, we have shown how quantum discord can feature genuine critical behavior even in cases where stricter forms of quantum correlations - most notably, entanglement - are very short-ranged. 
The phenomenology we describe does not rely on the system being at thermal equilibrium, so that it could be in principle adapted to search for signatures of quantum correlations in steady states of Liouville dynamics (where the role played by entanglement, if any, is unclear)~\cite{PhysRevLett.116.070407}, possibly in combination with complementary aspects such as quantum coherence~\cite{Roscilde2016,Fr_rot_2016}.

\paragraph{Acknowledgements. -} We thank R. Fazio, T. Grover, T-C. Lu and T. Roscilde for discussions and feedback on the manuscript. 
This work is partly supported by the ERC under grant number 758329 (AGEnTh), by the MIUR Programme FARE (MEPH), and by European Union's Horizon 2020 research and innovation programme under grant agreement No 817482 (Pasquans). P.S.T acknowledges support from the Simons Foundation Award 284558FY19 to the ICTP. F. I. acknowledges support by MUR (Ministero dell’Universit\'a e della Ricerca) via project PRIN 2017 "Taming complexity via QUantum Strategies: a Hybrid Integrated Photonic approach" (QUSHIP) Id. 2017SRNBRK.

{\it Note added:} while this manuscript was in preparation, we became aware of a recent work by Ma and Sela~\cite{Ma:2021aa} that also discusses persistence of quantum correlations at finite temperature.

\addcontentsline{toc}{chapter}{Bibliography} 

\bibliography{Discord_FI.bib}

\setcounter{equation}{0}
\setcounter{figure}{0}
\setcounter{table}{0}
\renewcommand{\theequation}{S\arabic{equation}}
\renewcommand{\thefigure}{S\arabic{figure}}

\clearpage
\onecolumngrid
\begin{center}
    \textbf{\Large Supplementary Materials}
\end{center}
\vspace*{1cm}

\twocolumngrid

\section{Covariance matrix}
Our model is described by Gaussian states, which are completely characterized by their covariance matrix. This fact allows one to compute various quantities efficiently, including the quantum discord. In particular, for our model, the elements of the covariance matrix (two-point correlators) are 
\be \label{eq:phi_correlator}
 \sigma_{\phi}(\vec{r}-\vec{r}')=2\langle \phi_{\vec{r}}  \phi_{\vec{r}'} \rangle=\frac{1}{N} \sum_{\vec{k}} e^{i\vec{k}\cdot \left(\vec{r} -\vec{r}'  \right)} \frac{1}{\omega_{\vec{k}}} \coth(\frac{1}{2} \beta \omega_{\vec{k}}) 
\ee
\be \label{eq:pi_correlator}
  \sigma_{\pi}(\vec{r}-\vec{r}')=2\langle \pi_{\vec{r}}  \pi_{\vec{r}'} \rangle=\frac{1}{N} \sum_{\vec{k}} e^{i\vec{k}\cdot \left(\vec{r} -\vec{r}'  \right)} \omega_{\vec{k}} \coth(\frac{1}{2} \beta \omega_{\vec{k}}) ,
\ee
where $\vec{k}=(k_1,k_2,\cdots,k_d)=\frac{2\pi}{L}(n_1,n_2,\cdots,n_d)$ for $n_i=0,1,\cdots,L-1$, and $\omega_{\vec{k}}=\sqrt{m^2+4K\sum_{i=1}^d \sin^2(\frac{k_i}{2})}$. Note that for finite-size systems, at the critical point, i.e. $m=0$, the contribution of the zero mode for $\sigma_{\phi}(\vec{r}-\vec{r}')$ diverges. On the other hand, in the thermodynamic limit, $\sigma_{\phi}(\vec{r}-\vec{r}')$ is finite for $d>2$ and diverges otherwise. Focusing on two modes at sites $\vec{r}$ and $\vec{r}'$ and tracing over all other modes, the covariance matrix reads
\be
\sigma=
\begin{pmatrix}
\sigma_{\phi}(0) & 0 & \sigma_{\phi}(\vec{r}-\vec{r}') & 0 \\  
0 & \sigma_{\pi}(0) & 0 & \sigma_{\pi}(\vec{r}-\vec{r}') \\
\sigma_{\phi}(\vec{r}-\vec{r}') & 0 & \sigma_{\phi}(0) & 0 \\
0 & \sigma_{\pi}(\vec{r}-\vec{r}') & 0 & \sigma_{\pi}(0) \\
\end{pmatrix}.
\ee
The previous formulation can then be brought in a more manageable form as discussed in the text.

\section{Geometric Discord}
Another way to measure the quantum discord is via a geometrical approach. This measure is called the {\it{geometric measure of quantum discord}}, or the geometric discord (GD) for short. It is defined as \cite{dakic2010},
\be
    D_G(\rho_{AB})=\frac{1}{N_d}\min_\sigma d(\rho,\sigma)^2,
\ee
where the minimization is over the set of states $\sigma$ with zero quantum discord (classical-quantum states), $d$ is a distance function on the set of quantum states, and $N_d$ is a normalization constant such that $D_G\in[0,1]$. It follows that the GD vanishes on the set of classical-quantum states, as is the case for the entropic discord, while the amount of quantum correlations of a given state is quantified by how "far" apart the state is from the set. 

Initially, the distance most commonly used has been the Hilbert-Schmidt distance $d_2(\rho,\sigma)=\[\tr(|\rho-\sigma|^2)\]^{1/2}$ \cite{dakic2010}. However, it turns out that the Hilbert-Schmidt distance is not contractive under quantum operations, a property that is necessary for physically reliable distances \cite{piani2012}. Thus, the GD based on the Hilbert-Schmidt distance is not a good measure of quantum correlations. On the other hand, prominent examples of contractive distances equipped with a physical and operational meaning are the Bures and Hellinger distances, defined as \cite{spehner2016}
\be
    d_{\textrm{Bu}}(\rho,\sigma)=\(2-2\sqrt{\mathcal{F}(\rho,\sigma)}\)^{1/2}
\ee
\be
    d_{\textrm{He}}(\rho,\sigma)=\(2-2\mathcal{A}(\rho,\sigma)\)^{1/2},
\ee
where the Uhlmann fidelity $\mathcal{F}(\rho,\sigma)$ and the affinity $\mathcal{A}(\rho,\sigma)$ are given by 
\be
    \mathcal{F}(\rho,\sigma)=\(\tr[(\sqrt{\sigma}\rho\sqrt{\sigma})^{1/2}]\)^{2},
\ee
\be
    \mathcal{A}(\rho,\sigma)=\tr\sqrt{\rho}\sqrt{\sigma},
\ee
respectively. Apart of being contractive, these two distances also enjoy some other desirable mathematical properties, making them the most prominent metrics used to quantify quantum correlations (see Ref. \cite{spehner2016} for a thorough review).

In the context of Gaussian states, the Gaussian geometric discord is defined accordingly as the minimum squared distance between a Gaussian state and the set of classical-quantum Gaussian states \cite{adesso2011_2}. However, the set of Gaussian states which are classical-quantum are known to consist of only product states. Therefore, the Gaussian geometric discord measures the total (classical and quantum) correlations, i.e., it cannot be the true geometric discord. Nevertheless, it is still an interesting quantity since it provides an upper bound to the true geometric discord and in terms of the Hellinger distance it is computable for all two-mode Gaussian states \cite{marian2015,spehner2016}. 

\begin{figure} [t]
    \centering
    \includegraphics[width=0.48\textwidth]{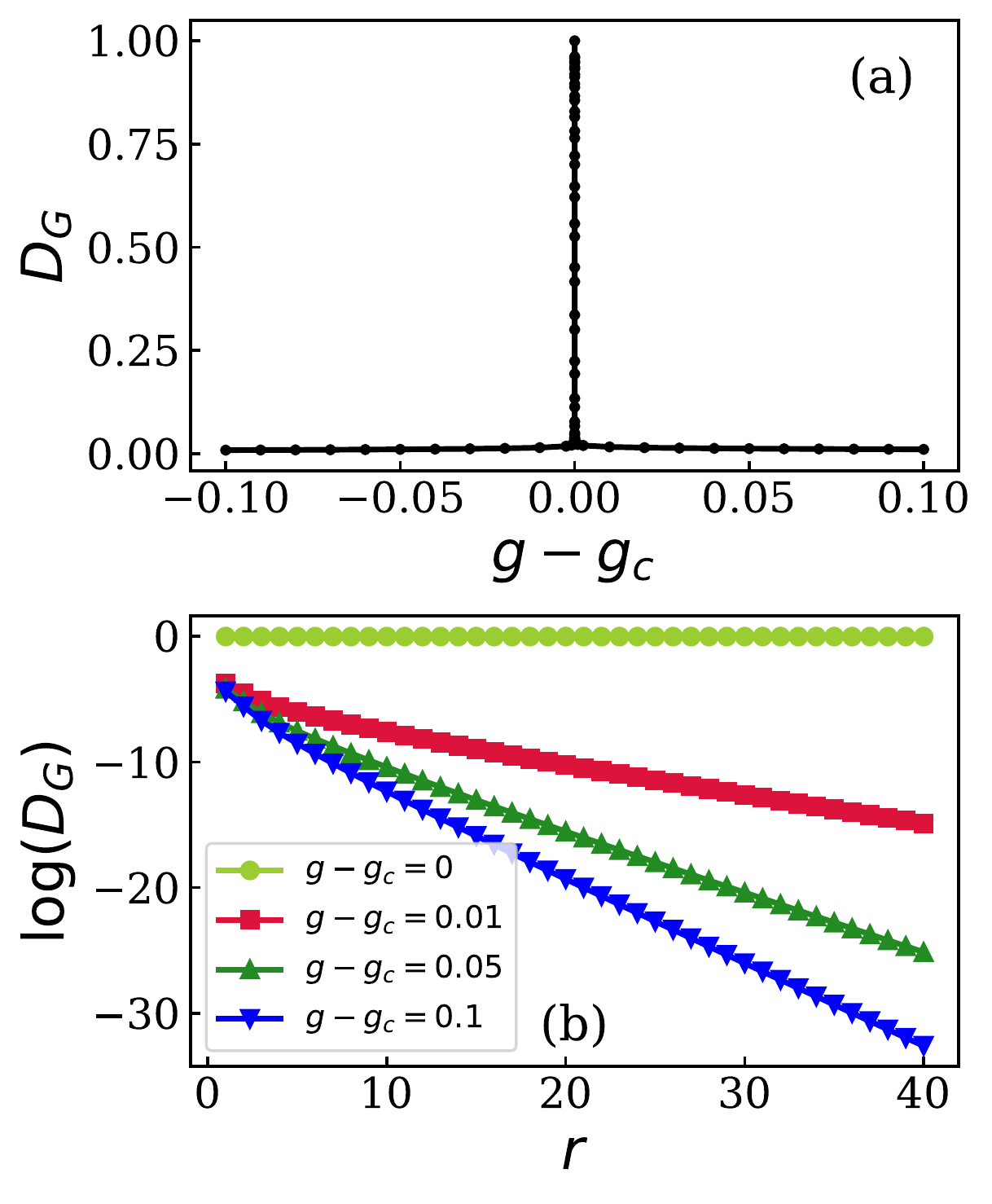}
    \caption{(a) Behavior of the geometric discord between nearest-neighbor pairs as a function of $g-g_c$ on a 2D square lattice. (b) Decay of geometric discord as a function of the distance.}
    \label{fig:gqd}
\end{figure}

We thus study the behavior of the GD based on the Hellinger metric on the same model considered in the main text across a finite-temperature phase transition. In Fig. (\ref{fig:gqd}a), we show the behavior of the GD for nearest-neighbors in a system of $N=10^6$ sites. Again, the local maximum is found at the critical point $g=g_c$, similarly to the EQD. Concerning the long-distance properties, Fig. (\ref{fig:gqd}b) shows the decay of the GD as a function of the distance for several values of $g$. We find that away from the critical point the GD vanishes exponentially. On the other hand, exactly at the critical point, the GD is uniformly one at any distance, which is the largest value $D_G$ can assume. Interestingly, we find that the discord length corresponding to the GD is identical to that of the EQD in our model.

\begin{figure} [t]
    \centering
    \includegraphics[width=0.48\textwidth]{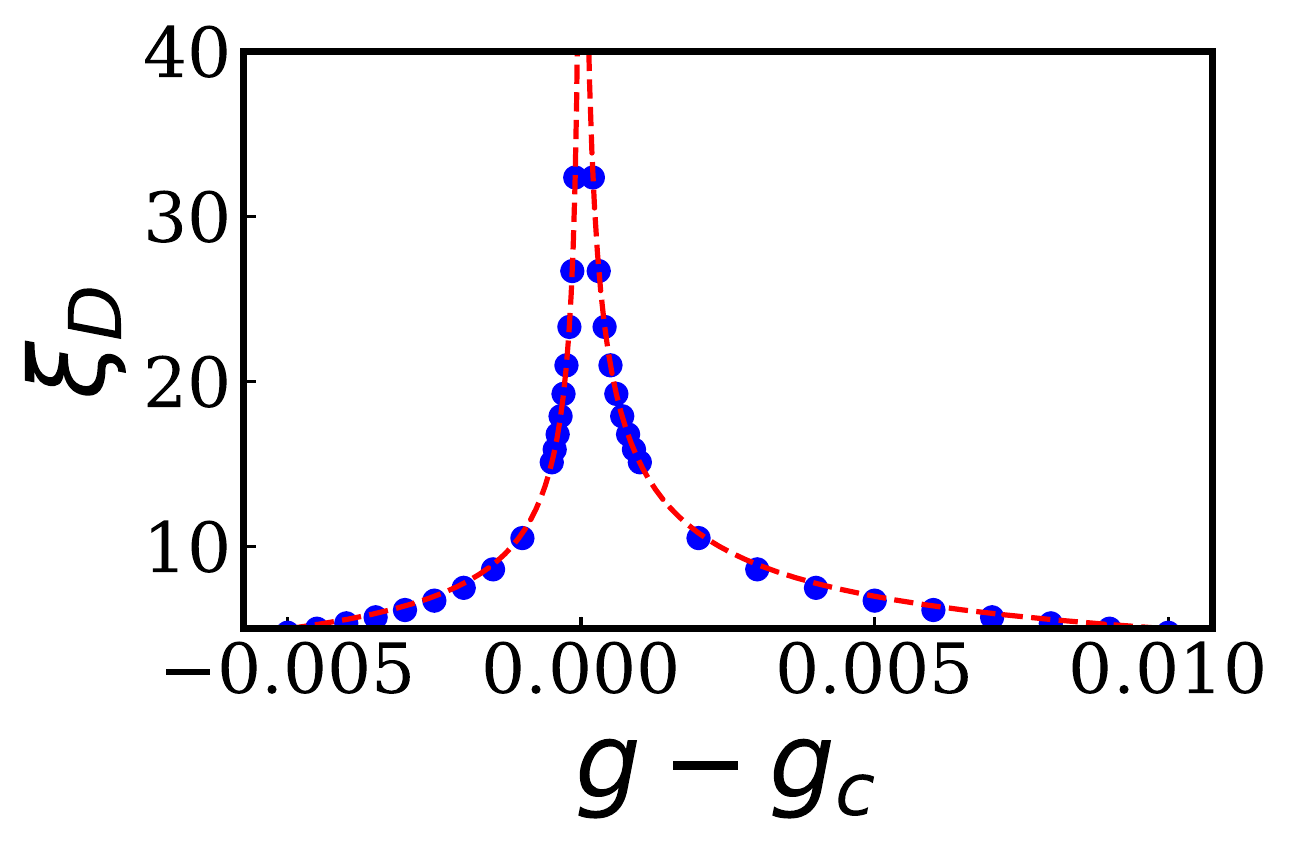}
    \caption{Behavior of the discord length near the critical point at $T=0.2$. The red dashed line is the power law fit $\xi_D=a_{\pm}|g-g_c|^{-\nu_D}$ where the exponent is
    $\nu_D=0.4785 \pm 0.0045$}
    \label{fig:discordlength}
\end{figure}

\section{Discord length and finite-size scaling}
\begin{figure}
    \centering
    \includegraphics[width=0.48\textwidth]{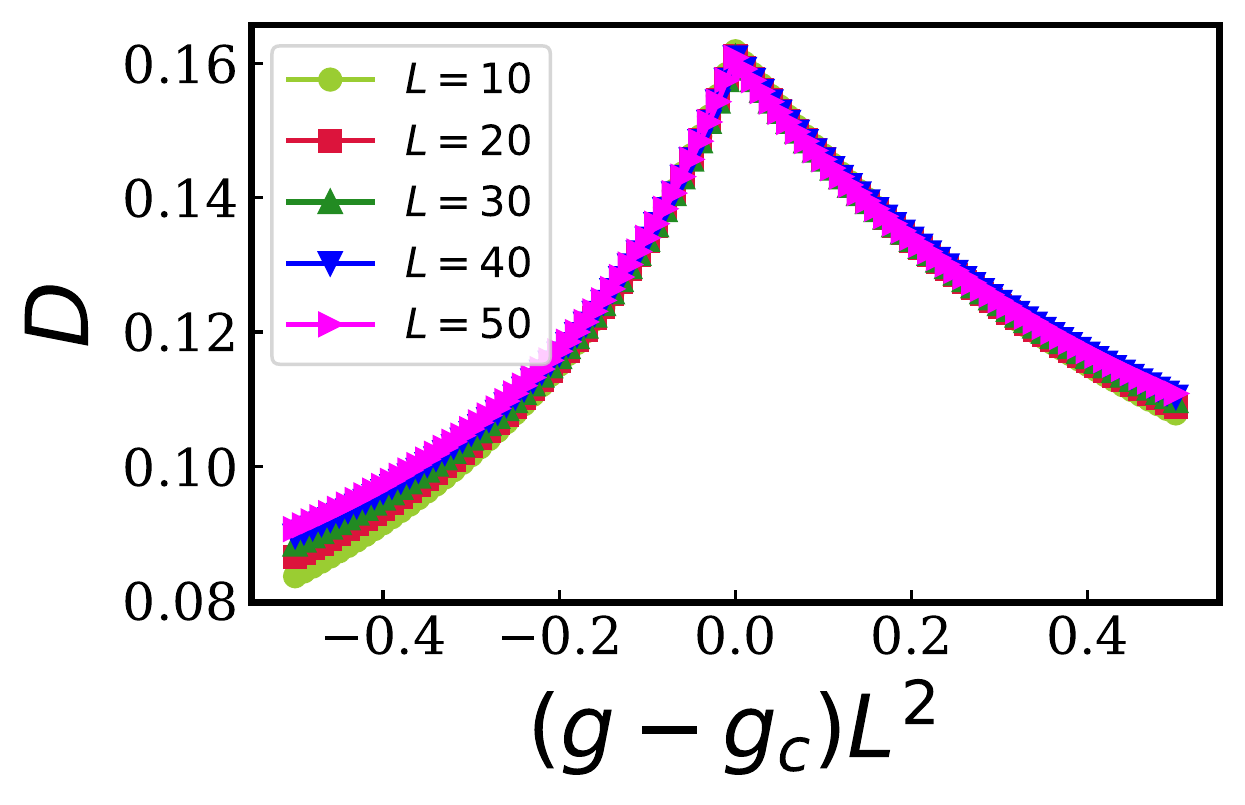}
    \caption{Data collapse of the entropic quantum discord EQD with respect to $(g-g_c)L^{1/\nu}$.}
    \label{fig:fss}
\end{figure}

 In Fig. \ref{fig:discordlength}, we show the behavior of the discord length $\xi_Q$ around the critical point $g=g_c$ at $T=0.2$. It is observed that $\xi_Q$ exhibits a power-law divergent behavior, $\xi_D\sim|t|^{-\nu_D}$, with $t=g-g_c$. The power-law exponent is found to be $\nu_D=0.4785 \pm 0.0045$. In Fig. \ref{fig:fss}, we show the finite-size scaling collapse of the nearest-neigbor EQD.

\section{3D lattice}
\begin{figure} [t]
    \centering
    \includegraphics[width=0.4\textwidth]{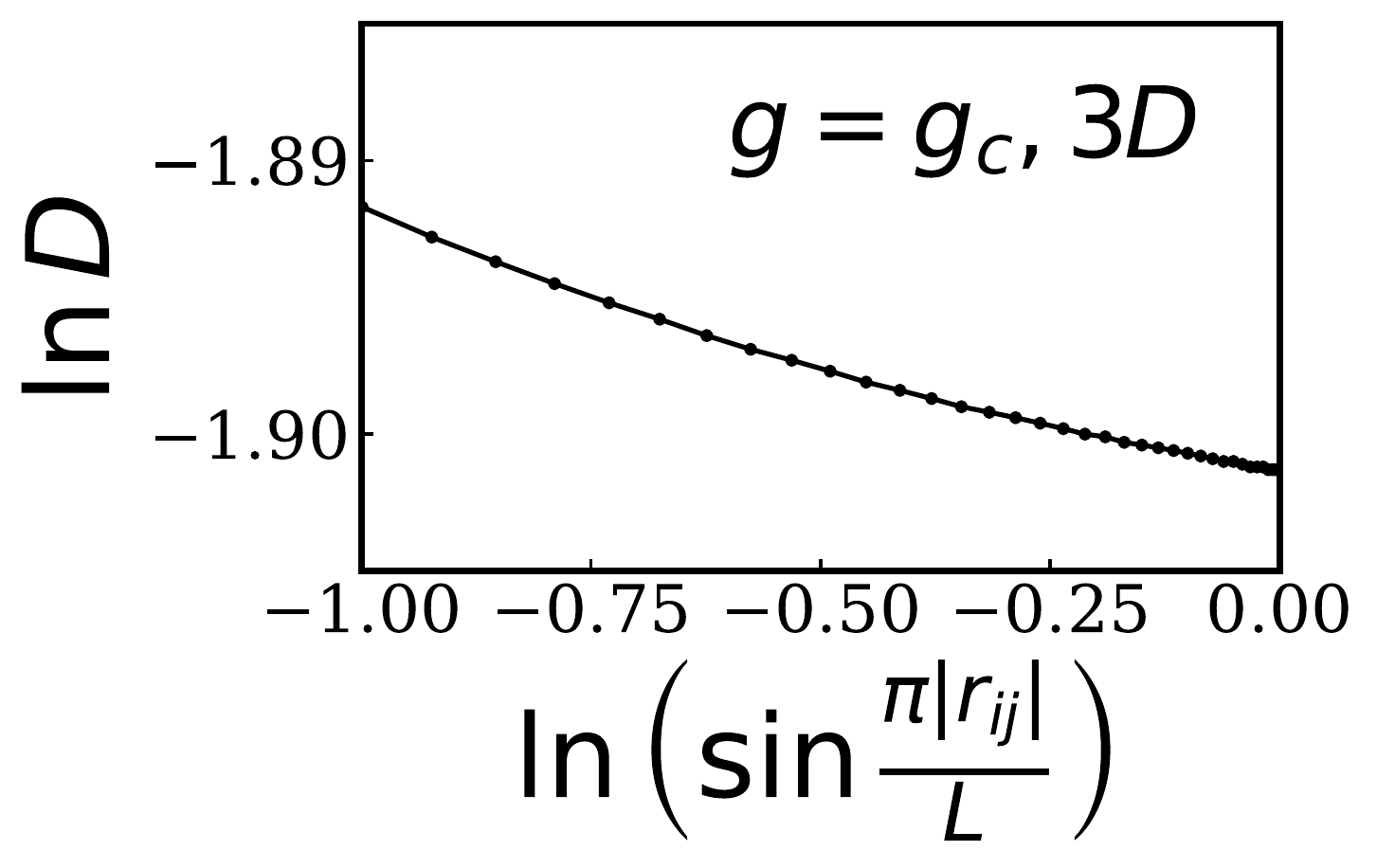}
    \caption{Decay of the entropic quantum discord EQD in 3D systems at the critical point.}
    \label{fig:qd3D}
\end{figure}

In the 3D case, we observe that the entropic quantum discord EQD displays qualitatively a very similar critical behavior as that displayed in the 2D case. For example, the discord length diverges at the critical point, whereas the EQD displays a power law decay, as shown in Fig. \ref{fig:qd3D}.

\end{document}